# THE SOLUTION OF COMPLEX PROBLEMS ON CALCULATION OF THE ELECTROSTATIC FIELDS ON LESSONS ON COMPUTER MODELING

Mayer Robert Valerievich, http://maier-rv.glazov.net
The Glazov Korolenko State Pedagogical Institute, Glazov, Russia

**Abstract:** In article the following tasks on computer modeling of electric fields are analyzed: 1) calculation of distribution of potential for the field created by two parallel plates and charged bodies in the non-uniform environment; 2) calculation of distribution of potential and force lines of electric field in which are brought the cylinder, a pipe, a plate, a rectangular parallelepiped from dielectric, and also the metal cylinder; 3) calculation of distribution of potential in the one-dimensional non-uniform environment; 4) the solution of the equation of Poisson in spherical coordinates; 5) calculation of distribution of potential in cylindrical coordinates with the subsequent creation of equipotential surfaces and force lines.

# РЕШЕНИЕ СЛОЖНЫХ ЗАДАЧ НА РАСЧЕТ ЭЛЕКТРОСТАТИЧЕСКОГО ПОЛЯ НА ЗАНЯТИЯХ ПО КОМПЬЮТЕРНОМУ МОДЕЛИРОВАНИЮ

Майер Роберт Валерьевич, http://maier-rv.glazov.net
Глазовский государственный педагогический институт, Глазов, Россия

При изучении основ компьютерного моделирования имеет смысл рассмотреть вопрос о численном решении уравнения Пуассона при расчете электростатического поля в однородный и неоднородных средах в различных системах координат. Допустим, известны плотность распределения заряда и потенциал на границе области двумерной неоднородной среды. В декартовых координатах распределение потенциала удовлетворяет уравнению:

$$\frac{\partial}{\partial x}\left(\varepsilon(x,y)\frac{\partial \varphi}{\partial x}\right) + \frac{\partial}{\partial y}\left(\varepsilon(x,y)\frac{\partial \varphi}{\partial y}\right) = -\frac{\rho(x,y)}{\varepsilon_0},$$

$$\frac{\partial \varepsilon}{\partial x}\frac{\partial \varphi}{\partial x} + \varepsilon\frac{\partial^2 \varphi}{\partial x^2} + \frac{\partial \varepsilon}{\partial y}\frac{\partial \varphi}{\partial y} + \varepsilon\frac{\partial^2 \varphi}{\partial y^2} = -\frac{\rho}{\varepsilon_0}.$$

В конечных разностях получаем:

$$\frac{\varepsilon_{i+1,j} - \varepsilon_{i-1,j}}{2h}\frac{\varphi_{i+1,j} - \varphi_{i-1,j}}{2h} + \varepsilon_{i,j}\frac{\varphi_{i-1,j} - 2\varphi_{i,j} + \varphi_{i+1,j}}{h^2} +$$

$$\frac{\varepsilon_{i,j+1}-\varepsilon_{i,j-1}}{2h}\frac{\varphi_{i,j+1}-\varphi_{i,j-1}}{2h}+\varepsilon_{i,j}\frac{\varphi_{i,j-1}-2\varphi_{i,j}+\varphi_{i,j+1}}{h^2}=-\frac{\rho_{i,j}}{\varepsilon_0},$$

$$\varphi_{i,j}=(\varphi_{i+1,j}+\varphi_{i-1,j}+\varphi_{i,j+1}+\varphi_{i,j-1}+\rho_{i,j}h^2/\varepsilon_0\varepsilon_{i,j})/4+[(\varepsilon_{i+1,j}-\varepsilon_{i-1,j})\cdot(\varphi_{i+1,j}-\varphi_{i-1,j})+(\varepsilon_{i,j+1}-\varepsilon_{i,j-1})\cdot(\varphi_{i,j+1}-\varphi_{i,j-1})]/(16\varepsilon_{i,j}).$$

Для решения рассматриваемой стационарной задачи используется релаксационный метод последовательных приближений: во всех внутренних узлах сетки задаются произвольные исходные значения потенциала; его значения во внешних узлах должны соответствовать граничным условиям. Осуществляется первая итерация, в ходе которой перебираются все внутренние узлы сетки и, исходя из начальных значений, определяют новые уточненные значения функции. Затем осуществляются второе, третье, … приближения, причем результаты i–ой итерации используются в качестве исходных для (i+1)-ой итерации. В результате получающиеся значения приближаются к истинному распределению потенциала.

Для реализации этого алгоритма требуется зарезервировать в памяти ЭВМ три двумерных массива размерностью N x M, элементами которых являются числа типа real или single (язык Pascal). Один массив для диэлектрической проницаемости, два других — для потенциалов на шаге t и t+1. Это не позволяет в среде Free Pascal создать сетку с большим числом узлов. Увеличить число узлов можно, задав диэлектрическую проницаемость целыми числами. Вместо двух массивов для потенциалов на различных временных слоях следует создать один, но в этом случае, чтобы избежать накапливания ошибок, придется перебирать элементы в четырех различных направлениях.

Алгоритм состоит в последовательном переборе узлов сетки слева на право и справа на лево, сверху вниз и снизу вверх, в ходе которого вычисляются значения потенциала в них. Перед каждым проходом следует учесть граничные условия задачи, то есть приравнять потенциалы на границе двумерной области к заданной величине или функции координат. Используемая программа ПР–1 содержит: 1) процедуру Sreda, в которой задается диэлектрическая проницаемость среды в узлах сетки; 2) процедуру Gran, в которой учитывается распределение потенциала вдоль границы области; 3) процедуру Raschet, в которой задается распределение заряда и вычисляется потенциал в различных узлах (i, j); 4) процедуру Draw, выводящую результат вычислений на экран; 5) основную часть программы, в которой осуществляется вызов перечисленных процедур в требуемом порядке.

На рис. 1.1 и 2 представлены результаты ее использования для расчета электрического поля в двумерной неоднородной среде с диэлектрической проницаемостью $\varepsilon = 1 + axy$, созданным двумя пластинами, которые имеют потенциалы 350 и –350 В, и зарядами $q_1$ и $q_2$. В пространстве между пластинами имеется заземленный проводник, его потенциал равен нулю.

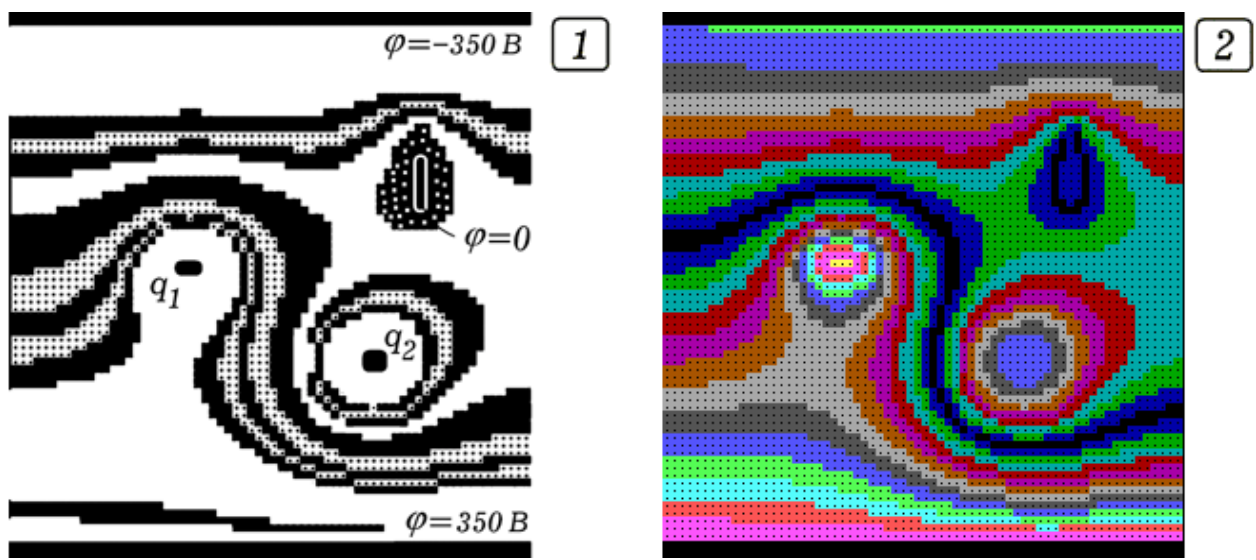

Рис. 1. Расчет потенциала электрического поля в неоднородной среде.

Задавая различные распределения зарядов, потенциалов, диэлектрической проницаемости среды и граничные условия, можно решить и другие задачи. Например, рассчитать электрическое поле, создаваемое двумя параллельными разноименно заряженными пластинами, в случаях, когда в пространство между ними внесен: 1) прямоугольный брусок с диэлектрической проницаемостью меньшей, чем у окружающей среды (рис. 2.1); 2) прямоугольный брусок с диэлектрической проницаемостью большей, чем у окружающей среды (рис. 2.2); 3) цилиндр с диэлектрической проницаемостью большей, чем у окружающей среды (рис. 2.3); 4) труба с с диэлектрической проницаемостью большей, чем у окружающей среды (рис. 2.4); 5) пластина с диэлектрической проницаемостью большей, чем у окружающей среды, расположенная под углом (рис. 2.5); 6) металлический цилиндр (рис. 2.6). Программа ПР–2 рассчитывает линии равного потенциала и силовые линии электрического поля в случае, когда в пространство между заряженными пластинами внесена труба с диэлектрической проницаемостью большей, чем у окружающей среды. Из рисунков видно, что на границе раздела различных сред происходит преломление силовых линий; причем в средах с большей диэлектрической проницаемостью они располагаются реже (напряженность меньше). При внесении в электрическое поле проводящего цилиндра (потенциалы всех его точек равны), силовые линии оказываются перпендикулярными к поверхности.

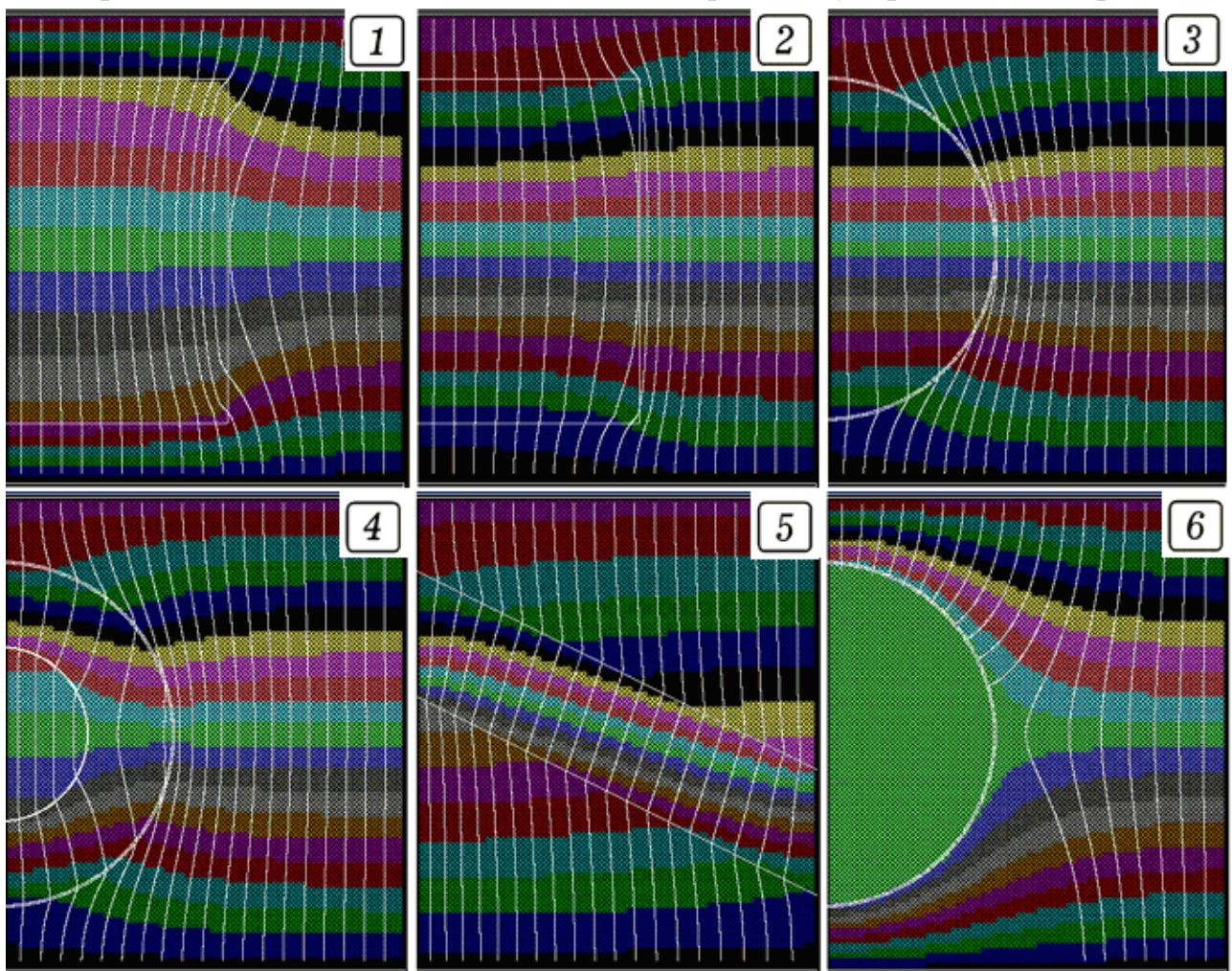

Рис. 2. Результаты расчета электростатического поля.

Рассмотренный выше подход позволяет решить одномерную задачу, например, рассчитать распределение потенциала электростатического поля между двумя бесконечно большими пластинами 1 и 3, расположенными перпендикулярно оси Ox на расстоянии b друг от друга (рис. 3.1). Потенциалы пластин заданы; между ними расположена пластина 2 из диэлектрика, заряженная положительно или отрицательно. Рядом с пластиной 3 имеется слой диэлектрика с некоторой диэлектрической проницаемостью. На рис. 3.2 показано получающееся распределение потенциала вдоль оси Ox в случае, когда между металлическими пластинами имеется толстая пластина с заданной диэлектрической проницаемостью, внутри которой равномерно "размазан" электрический заряд. Представлено не только искомое распределение потенциала, но и несколько предшествующих ему приближений; это помогает понять сущность метода последовательных итераций.

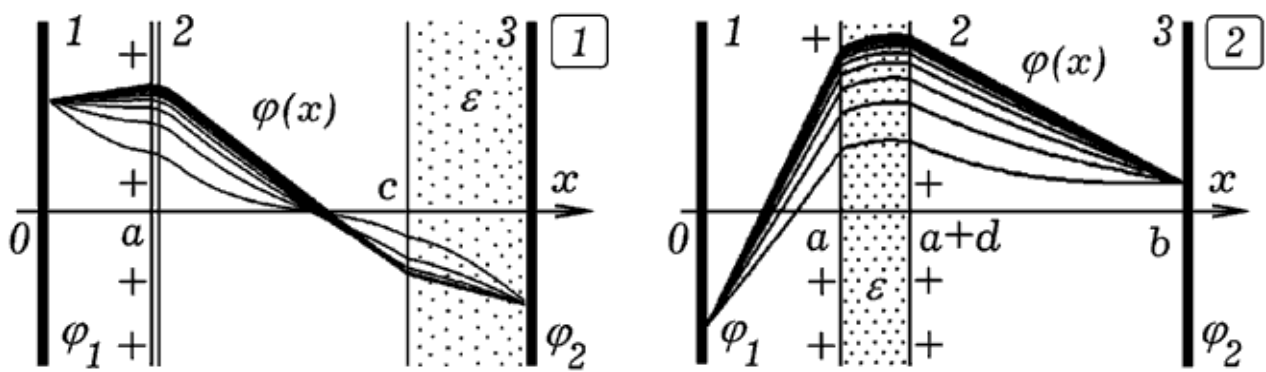

Рис. 3. Расчет потенциала поля в одномерном случае.

Выше рассмотрены методы решения уравнения Пуассона в декартовых координатах. Теперь рассчитаем распределение потенциала в двумерной области, имеющей форму четверти круга, для чего решим уравнение Пуассона в полярных координатах (рис. 4.1). Граничные условия зададим так:

1) граница AB: $(d\varphi/d\alpha)_{AB} = 0$, $\varphi_{i,0} = \varphi_{i,1}$;

2) граница BC: $\varphi_{BC} = 20$ В, $\varphi_{N,j} = 20$ В;

3) граница CD: $(d\varphi/d\alpha)_{CD} = 0$, $\varphi_{i,M+1} = \varphi_{i,M}$;

4) граница DA: $(d\varphi/d\alpha)_{DA} = 0$, $\varphi_{0,j} = 0$ В.

Потенциалы двух точек поддерживаются равными 50 В и – 80 В. Уравнение Пуассона в полярных координатах и соответствующее конечно-разностное уравнение имеют вид:

$$\frac{1}{r}\frac{\partial \varphi}{\partial r} + \frac{\partial^2 \varphi}{\partial r^2} + \frac{1}{r^2}\frac{\partial^2 \varphi}{\partial \alpha^2} = -\frac{q(x,y)}{\varepsilon\varepsilon_0},$$

$$\frac{\varphi_{i+1,j} - \varphi_{i-1,j}}{2r\Delta r} + \frac{\varphi_{i+1,j} - 2\varphi_{i,j} + \varphi_{i-1,j}}{\Delta r^2} + \frac{\varphi_{i,j+1} - 2\varphi_{i,j} + \varphi_{i,j-1}}{r^2\Delta\alpha^2} = -\frac{q_i}{\varepsilon\varepsilon_0},$$

$$2\left(\frac{1}{r^2\Delta\alpha^2} + \frac{1}{\Delta r^2}\right)\varphi_{i,j} = \frac{\varphi_{i+1,j} - \varphi_{i-1,j}}{2r\Delta r} + \frac{\varphi_{i+1,j} + \varphi_{i-1,j}}{\Delta r^2} +$$

$$+ \frac{\varphi_{i,j-1} + \varphi_{i,j+1}}{r^2\Delta\alpha^2} + \frac{q_{i,j}}{\varepsilon\varepsilon_0}.$$

Отсюда выражают потенциал в узле (i, j). Для решения задачи используется программа ПР-3. Результаты вычисления распределения потенциала в двумерной области при заданных граничных условиях представлены на рис. 4.2. Программа выполняет последовательность итераций, получающиеся значения потенциалов постепенно приближаются к искомым значениям, которые соответствуют точному решению задачи.

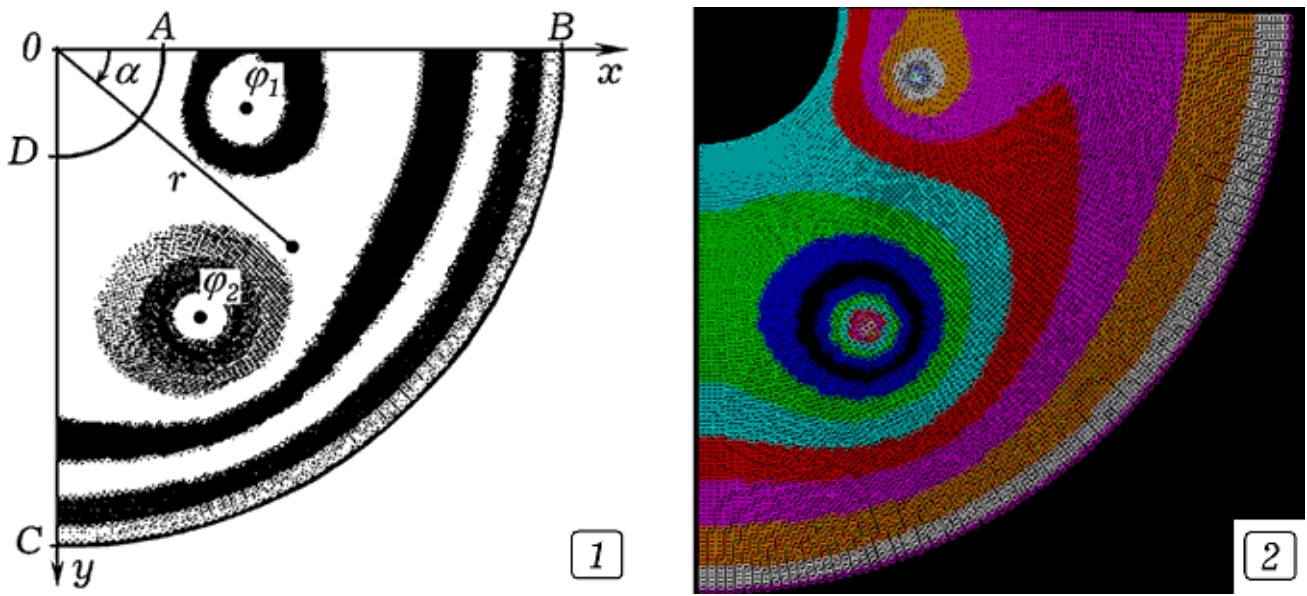

Рис. 4. Расчет электрического поля в сферических координатах.

Теперь рассчитаем электростатическое поле в пространстве между двумя электродами 1 и 2 при наличии заряженного шара 3 (рис. 5.1), используя цилиндрические координаты. Распределение потенциала должно удовлетворять уравнению Пуассона:

$$\frac{1}{r}\frac{\partial \varphi}{\partial r} + \frac{\partial^2 \varphi}{\partial r^2} + \frac{1}{r^2}\frac{\partial^2 \varphi}{\partial \alpha^2} + \frac{\partial^2 \varphi}{\partial z^2} = -\frac{\rho}{\varepsilon\varepsilon_0}.$$

Электроды обладают цилиндрической симметрией, то же самое относится к распределению заряда и граничным условиям. Задача сводится к нахождению распределения потенциала в плоскости осевого сечения xOz. В конечных разностях получаем:

$$\frac{\varphi_{i+1,k} - \varphi_{i-1,k}}{2r\Delta r} + \frac{\varphi_{i-1,k} - 2\varphi_{i,k} + \varphi_{i+1,k}}{\Delta r^2} + \frac{\varphi_{i,k-1} - 2\varphi_{i,k} + \varphi_{i,k+1}}{\Delta z^2} = -\frac{\rho_{i,k}}{\varepsilon\varepsilon_0},$$

$$\varphi_{i,k} = \frac{\varphi_{i+1,k} - \varphi_{i-1,k}}{8r}h + \frac{\varphi_{i-1,k} + \varphi_{i+1,k} + \varphi_{i,k-1} + \varphi_{i,k+1}}{4} + \frac{\rho_{i,k}}{4\varepsilon\varepsilon_0}h^2.$$

Используется программа ПР–4, в которой перебираются все узлы двумерной сетки и с каждой итерацией пересчитываются значения потенциала. После 5000 итераций осуществляется построение силовых линий. Для этого в пространство между электродами запускаются невесомые частицы–маркеры, которые перемещаются вдоль силовых линий и рисуют свою траекторию. Напряженность поля равна градиенту потенциала. Для определения направления силовой линии в точке с координатами $x$, $z$ используются формулы:

$$E_x = -\frac{1}{2}\left(\frac{\varphi_{i+1,k} - \varphi_{i,k}}{h} + \frac{\varphi_{i+1,k+1} - \varphi_{i,k+1}}{h}\right),$$

$$E_z = -\frac{1}{2}\left(\frac{\varphi_{i,k+1} - \varphi_{i,k}}{h} + \frac{\varphi_{i+1,k+1} - \varphi_{i+1,k}}{h}\right).$$

Результаты расчета электростатического поля представлены на рис. 5.2. Силовые линии перпендикулярны эквипотенциальным поверхностям и направлены от положительного электрода к отрицательному.

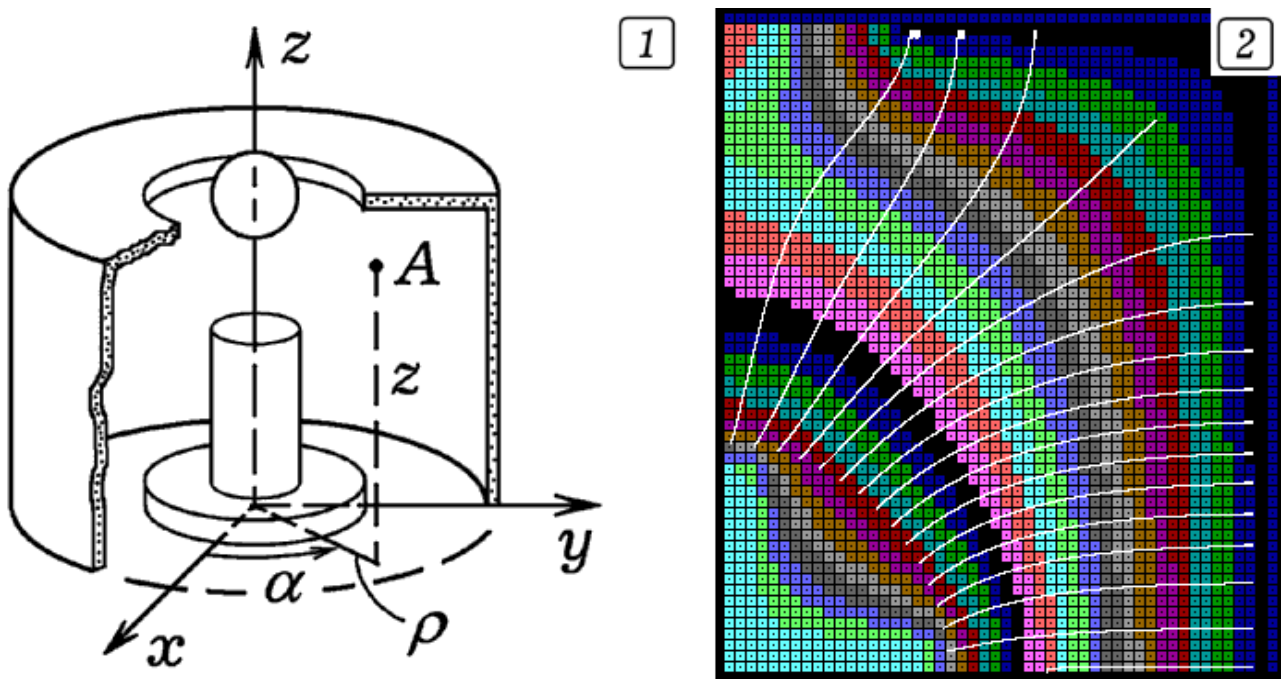

Рис. 5. Расчет электрического поля в цилиндрических координатах.

Обсуждение и анализ рассмотренных выше задач на занятиях по компьютерному моделированию способствуют более глубокому пониманию методов решения дифференциальных уравнений с частными производными, повышает интерес студентов к программированию, физике и математике.

Программа ПР–1.

```
program el_pole_neodnorodn_sreda;
{$N+} uses crt, graph; const N=70; M=70;
var i, j, ii, jj, k, q, DV, MV :integer;
fi, e: array[1..N, 1..M] of single;
Procedure Sreda;
begin For i:=1 to N do For j:=1 to M do e[i,j]:=1+0.015*j*i; end;
Procedure Gran;
begin
For i:=1 to N do begin fi[i,1]:=-350; fi[i,M]:=350;
  fi[1,i]:=fi[2,i]; fi[N,i]:=fi[N-1,i]; end;
For j:=15 to 25 do fi[55,j]:=0; end;
Procedure Raschet;
```

```
begin q:=0;
If ((i>20)and(i<30)and(j>30)and(j<35))then q:=500;
If ((i>40)and(i<55)and(j>40)and(j<55))then q:=-300;
fi[i,j]:=(fi[i+1,j]+fi[i-1,j]+fi[i,j+1]+fi[i,j-1]+q/
e[i,j])/4+((e[i-1,j]-e[i+1,j])*(fi[i+1,j]-fi[i-1,j]) +(e[i,j-1]-
e[i,j+1])*(fi[i,j+1]-fi[i,j-1]))/
(16*e[i,j]); end;
Procedure Draw;
begin setcolor(round(abs(2+fi[i,j]/30)));
   rectangle(i*4+50,j*4,i*4+54,j*4+4);
   rectangle(i*4+51,j*4+1,i*4+53,j*4+3); end;
BEGIN  DV:=Detect; InitGraph(DV,MV,'c:\bp\bgi'); Sreda;
Repeat inc(k);
   Gran; For i:=2 to N-1 do For j:=2 to M-1 do Raschet;
   Gran; For j:=M-1 downto 2 do
            For i:=N-1 downto 2 do Raschet;
   If k mod 50=0 then
      For i:=2 to N-1 do For j:=2 to M-1 do Draw;
until KeyPressed; CloseGraph;
END.
```

Программа ПР–2.

```
{$N+} program truba_el_pole;
uses crt, graph; const N=80; M=95; R=34; dt=0.1; h=1;
var i,j,q,s,DV,MV: integer; x,y,Ex,Ey:real;
fi,e: array[0..N+1,0..M+1] of single;
k: Longint; u: boolean;
Procedure Sreda;
begin For i:=1 to N do For j:=1 to M do begin
   u:=(sqr(i)+sqr(j-48)<R*R)and(sqr(i)+sqr(j-48)>R*R/4);
   {u:=(i<45)and(abs(j-48)<R);}
   If u then e[i,j]:=3 else e[i,j]:=1; end; end;
Procedure Gran;
begin For i:=1 to N do begin fi[i,1]:=100;
   fi[i,M]:=-100; end; For j:=1 to M do begin
   fi[1,j]:=fi[2,j]; fi[N,j]:=fi[N-1,j]; end; end;
Procedure Raschet;
begin fi[i,j]:=(fi[i+1,j]+fi[i-1,j]+fi[i,j+1]+fi[i,
j-1]+q/e[i,j])/4+((e[i-1,j]-e[i+1,j])*(fi[i+1,j]-fi
[i-1,j])+(e[i,j-1]-e[i,j+1])*(fi[i,j+1]-fi[i,j-1]))
/(16*e[i,j]); end;
BEGIN DV:=Detect; InitGraph(DV,MV,'c:\bp\bgi');
Sreda; x:=20; y:=10;
Repeat inc(k);
If k<2E+3 then begin
Gran; For i:=2 to N-1 do For j:=2 to M-1 do Raschet;
Gran; For j:=M-1 downto 2 do
         For i:=N-1 downto 2 do Raschet;
If k mod 200=0 then begin
   For i:=2 to N-1 do For j:=2 to M-1 do
   begin setcolor(round(10+fi[i,j]/10)mod 15);
   circle(50+i*5,j*5,2); circle(50+i*5,j*5,1); end;
end; end;
```

```
If (k>2E+3)and(x<5*80) then begin
If (y>5*93) then begin inc(s); x:=20*s; y:=10; end;
i:=round(int(x/5)); j:=round(int(y/5));
Ex:=-(fi[i+1,j]-fi[i,j]+fi[i+1,j+1]-fi[i,j+1])/h/2;
Ey:=-(fi[i,j+1]-fi[i,j]+fi[i+1,j+1]-fi[i+1,j])/h/2;
x:=x+Ex*dt; y:=y+Ey*dt;
setcolor(white); circle(50+round(x),round(y),1); end;
If k=2E+3 then begin
   setcolor(white); circle(50,48*5,R*5);
   circle(50,48*5,round(R*5/2+2)); end;
until KeyPressed; CloseGraph;
END.
```

Программа ПР–3.

```
Program el_pole_polyarnie_koordinat;
uses crt, graph;
const N=100; M=100; a=0.4; R0=3; dr=0.1; dt=0.005;
var i,j,k,Gd,Gm: integer; q,dal,al,pi,r: real;
fi: array[0..N+1,0..M+1] of real;
BEGIN
Gd:=Detect; InitGraph(Gd, Gm, 'c:\bp\bgi');
pi:=arctan(1)*4; dal:=pi/M/2;
Repeat
  For i:=1 to N do For j:=1 to M do begin r:=R0+i*dr;
    fi[20,20]:=50; fi[50,70]:=-80; fi[N,j]:=20;
    fi[i,j]:=((fi[i+1,j]-fi[i-1,j])/2/r/dr+(fi[i,j+1]+
     fi[i,j-1])/r/r/dal/dal+(fi[i-1,j]+fi[i+1,j])/dr/dr
     +q)/2/(1/r/r/dal/dal+1/dr/dr);
    fi[i,0]:=fi[i,1]; fi[i,M+1]:=fi[i,M]; end;
  If k mod 10=0 then For i:=1 to N do For j:=1 to M do
    begin r:=30*(R0+i*dr); al:=dal*j;
      setcolor(round(abs(fi[i,j]/5+3))); circle(100+
        round(r*cos(al)),20+round(r*sin(al)),2);
      end; inc(k);
until KeyPressed; CloseGraph;
END.
```

Программа ПР–4.

```
{$N+}program Potenc_cilindr_koordinati;
uses crt, graph;
const N=50; M=60; h=1; dt=0.01;
var i,j,k,DV,MV: integer; r,C,Ex,Ey,x,y : single;
t:Longint; f,f1: array[0..N+1,0..M+1] of single;
procedure Raschet;
begin r:=h*i+0.001;
If i*i+(M-k)*(M-k)<64 then C:=1 else C:=0;
f1[i,k]:=(f[i-1,k]+f[i+1,k]+f[i,k-1]+f[i,k+1]+(f[i+1,k]-f[i-
1,k])*h/2/r+C)/4; end;
BEGIN DV:=Detect; InitGraph(DV,MV,'c:\bp\bgi'); x:=30;
Repeat inc(t); If t<5000 then begin
For i:=0 to N do For k:=0 to M do begin
If (i<4)and(k<20)then f[i,k]:=30;
If (i<20)and(k<4)then f[i,k]:=30;
```

```
If (i>N-2)or((k>M-2)and(i>15)) then f[i,k]:=-30;
f[i,M]:=f[i,M-1]; f[i,0]:=f[i,1]; f[0,k]:=f[1,k]; end;
For i:=1 to N-1 do For k:=1 to M-1 do Raschet;
For i:=0 to N do For k:=0 to M do f[i,k]:=f1[i,k]; end;
If (t>5000)and(j<15) then begin If f[i,k]<-29.5 then
begin inc(j); x:=21*cos(j/10); y:=21*sin(j/10); t:=5000; end;
i:=round(x); k:=round(y);
Ex:=-(f[i+1,k]-f[i,k]+f[i+1,k+1]-f[i,k+1])/2;
Ey:=-(f[i,k+1]-f[i,k]+f[i+1,k+1]-f[i+1,k])/2;
x:=x+Ex*dt; y:=y+Ey*dt; setcolor(white);
circle(300+round(6*x),400-round(6*y),1); end;
If (t<5000)and(t mod 100=0) then begin
For i:=1 to N do For k:=1 to M do begin
setcolor(abs(round((f[i,k]+30)/2)mod 14));
rectangle(300+6*i,400-6*k,304+6*i,404-6*k);
rectangle(301+6*i,401-6*k,303+6*i,403-6*k); end; end;
until KeyPressed; CloseGraph;
END.
```

# Майер Р.В.
## Решение сложных задач на расчет электростатического поля на занятиях по компьютерному моделированию

В статье проанализированы следующие задачи по компьютерному моделированию электрических полей: 1) расчет распределения потенциала для поля, созданного двумя параллельными пластинами и заряженными телами в неоднородной среде; 2) расчет распределения потенциала и построение силовых линий электрического поля, в которое внесены цилиндр, труба, пластина, прямоугольный параллелепипед из диэлектрика, а также металлический цилиндр; 3) расчет распределения потенциала в одномерной неоднородной среде; 4) решение уравнения Пуассона в сферических координатах; 5) расчет распределения потенциала в цилиндрических координатах с последующим построением эквипотенциальных поверхностей и силовых линий.